\documentclass[onecolumn]{revtex4}
\usepackage{amssymb}

\def\al{\alpha}
\def\be{\beta}
\def\ga{\gamma}
\def\de{\delta}

\def\et{\eta}

\def\ka{\kappa}
\def\la{\lambda}

\def\si{\sigma}

\def\ph{\phi}

\def\ch{\chi}
\def\ps{\psi}
\def\om{\omega}

\def\La{\Lambda}

\def\fr#1#2{{{#1} \over {#2}}}
\def\half{{\textstyle{1\over 2}}}

\def\frac#1#2{{\textstyle{{#1}\over {#2}}}}

\def\lsim{\mathrel{\rlap{\lower4pt\hbox{\hskip1pt$\sim$}}
    \raise1pt\hbox{$<$}}}
\def\gsim{\mathrel{\rlap{\lower4pt\hbox{\hskip1pt$\sim$}}
    \raise1pt\hbox{$>$}}}
\def\sqr#1#2{{\vcenter{\vbox{\hrule height.#2pt
         \hbox{\vrule width.#2pt height#1pt \kern#1pt
         \vrule width.#2pt}
         \hrule height.#2pt}}}}

\def\pt#1{\phantom{#1}}

\def\nsc#1#2#3{\om_{#1}^{{\pt{#1}}#2#3}}

\def\vb#1#2{e_{#1}^{{\pt{#1}}#2}}
\def\ivb#1#2{e^{#1}_{{\pt{#1}}#2}}
\def\uvb#1#2{e^{#1#2}}

\def\barvb#1#2{\bar e_{#1}^{{\pt{#1}}#2}}

\newcommand{\beq}{\begin{equation}}
\newcommand{\eeq}{\end{equation}}
\newcommand{\bea}{\begin{eqnarray}}
\newcommand{\eea}{\end{eqnarray}}
\newcommand{\bit}{\begin{itemize}}
\newcommand{\eit}{\end{itemize}}
\newcommand{\rf}[1]{(\ref{#1})}

\begin{document}

\title{Gravity with Explicit Diffeomorphism Breaking}

\author{Robert Bluhm and Yumu Yang}

\affiliation{
Physics Department, Colby College,
Waterville, ME 04901  USA
}

\begin{abstract}
Modified  theories of gravity that explicitly break diffeomorphism invariance have been used for over a decade to explore open issues related to quantum gravity, dark energy, and dark matter.  At the same time, the Standard-Model Extension (SME) has been widely used as a phenomenological framework in investigations of spacetime symmetry breaking.  Until recently, it was thought that the SME was suitable only for theories with spontaneous spacetime symmetry breaking due to consistency conditions stemming from the Bianchi identities.  However, it has recently been shown that, particularly with matter couplings included, the consistency conditions can also be satisfied in theories with explicit breaking.  An overview of how this is achieved is presented, and two examples are examined.  The first is massive gravity, which includes a nondynamical background tensor.  The second is a model based on a low-energy limit of Ho\v rava gravity, where spacetime has a physically preferred foliation. In both cases, bounds on matter--gravity interactions that explicitly break diffeomorphisms are obtained using the SME.\end{abstract}

\maketitle

\section{Introduction}

Diffeomorphism invariance is a fundamental symmetry in Einstein's General Relativity (GR).
As a result of this symmetry and the Bianchi identities, 
when matter and gravity interact in GR,
the energy-momentum of the matter fields
is guaranteed to be covariantly conserved  
when the matter fields obey their equations of motion.  
Mathematically, this linkage between matter dynamics, geometrical constraints, and~energy-momentum conservation
is a result of Noether identities derived from the local diffeomorphism invariance of the Einstein--Hilbert and matter action~\cite{en18,Traut62}. 

As elegant and experimentally accurate as GR is as a classical theory,
it is widely believed that it must ultimately merge with quantum physics
at a very high-energy scale, such as the Planck scale.
It is hoped that in the context of a new quantum theory of gravity,
which would supersede GR,
unsolved problems having to do with quantization might find solutions and
unknown entities such as dark energy and dark matter might find physical explanations.
Amongst the many avenues of research looking for clues of a new quantum theory of gravity,
a number of them involve breaking local diffeomorphism invariance.
Examples include mechanisms in string field theory~\cite{ks1,ks2,ks3}, 
massive gravity~\cite{mg1,mg2}, 
Chern--Simons gravity~\cite{rjsp}, 
Bumblebee~\cite{bb1,objp05,bb2,rbngrpav,ms,ch14,amnp19,zo20,rmjn21} and Einstein--Aether models~\cite{ee07}, 
and Ho\v rava gravity~\cite{ph09,Muk10,Sot11,bps1,Wang17}.
The Standard-Model Extension (SME) provides a phenomenological theoretical framework
used in experimental tests of gravity and matter interactions that break spacetime symmetries~\mbox{\cite{sme1,sme2,sme3,akgrav04,rbsme,JT14,RB14,hees2016}},
and in recent decades, many high-precision bounds on possible symmetry-breaking terms have been obtained~\cite{aknr-tables}.

Diffeomorphism breaking can occur either spontaneously or explicitly.
In the case of spontaneous breaking, a~dynamical tensor field acquires
a nonzero vacuum expectation value,
which breaks the symmetry, or~more precisely hides the symmetry.
In this case, since all of the fields are dynamical, 
and excitations in the form of Nambu--Goldstone (NG) or Higgs-like excitations occur,
the usual linkage between matter dynamics, the~Bianchi identities, and~
covariant energy-momentum conservation still holds.
On the other hand, with~explicit breaking,
nondynamical backgrounds can appear or the action can contain terms
that are not scalars under the full diffeomorphism group.  
With explicit breaking,
the usual relations between dynamics, geometric identities, and~covariant energy-momentum conservation get disrupted,
and viable theories must ensure that there are no inconsistencies
that would rule them out.
To avoid this possibility, 
the gravity sector of the SME was originally constructed assuming
that the symmetry breaking was spontaneous~\cite{akgrav04}.

However, more recently, the question of whether inconsistencies
must occur or can be evaded when the symmetry breaking is explicit
has been examined more closely~\cite{rb15a,rb15b,rbas16,rbSym17,ybcc18,rbhbyw19,ybcp20,qb21,akzl21}.
It turns out that in many cases, the SME can still be applied to
gravity theories with explicit diffeomorphism breaking,
in particular when matter--gravity couplings are included.
At the same time, in~more restrictive limits, such as pure gravity in
a first-order perturbative treatment, evading inconsistency puts
very restrictive constraints on the geometry or the background fields,
which makes using the SME not so useful in these limits. 
One goal of this paper is to give an overview of how this plays out.
First, some general features are described.
Then,  specific examples of gravity models with explicit diffeomorphism breaking 
are examined, including massive gravity and a model
based on a low-energy or infrared (IR) limit of Ho\v rava gravity.
Matter--gravity couplings involving scalar and tensor
fields as well as fermions are investigated,
and by matching them to terms in the SME, experimental bounds
are~obtained.

It should be pointed out that recently, the gravity sector
of the SME has been generalized to include a much wider class 
of background fields at the level of effective field theory,
including backgrounds that explicitly break spacetime symmetry~\cite{akzl21}.
Many new phenomenological tests are likely to be performed 
using this broader set of interaction terms.
However, in~this paper, the question under consideration is 
whether previous tests of spacetime symmetry breaking that placed bounds on
terms in the original formulation of the SME (based on spontaneous
breaking) can be applied to the case of explicit breaking.  
The answer is that, in some cases, they can be when matter-gravity couplings are 
included, as can be shown using a St\"uckelberg~approach.

The outline of this paper is as follows.
To begin, some background on explicit diffeomorphism breaking is given in Section~\ref{sec2}.
This is followed in Section~\ref{sec3} by an examination of the consistency conditions that 
must hold with explicit breaking.
Section~\ref{sec4} looks at the question of whether the SME based on
spontaneous breaking can be used
to investigate gravity theories with explicit breaking.
Examples of how it can be used when matter-gravity couplings are 
included are given in Section~\ref{sec5}.
A summary and concluding remarks are then presented in Section~\ref{sec6}.

%%%%%%%%%%%%%%%%%%%%%%%%%%%%%%%%%%%%%%%%%%
\section{Explicit Diffeomorphism~Breaking} \label{sec2}

Infinitesimal diffeorphisms in GR are mappings of a spacetime manifold back to itself,
where points on the manifold are locally translated along a small vector $\xi^\mu$,
such that $x^\mu \rightarrow x^\mu + \xi^\mu$.
Changes in vector and tensor fields at a specified point $x^\mu$ are given by their Lie derivatives,
and collectively the transformations act like a gauge symmetry in GR.
For example, the~metric tensor changes as
\beq
g_{\mu\nu} \rightarrow g_{\mu\nu} + {\cal L}_\xi g_{\mu\nu} ,
\label{gLie}
\eeq
where ${\cal L}_\xi g_{\mu\nu} = D_\mu \xi_\nu + D_\nu \xi_\mu$
is the Lie derivative defined using covariant derivatives $D_\mu$.
The convention used in this paper is that the metric 
has signs $(-,+,+,+)$ on its diagonal, 
and it is assumed that there is no torsion.
Hence, the~covariant derivative of a spacetime tensor is defined
with a connection given by the Christoffel symbol.
A generic dynamical matter field,  denoted here as $f^\mu$, transforms as
\beq
f^\mu \rightarrow f^\mu + {\cal L}_\xi f^\mu ,
\label{fLie}
\eeq
where for a contravariant vector, ${\cal L}_\xi f^\mu = -(D_\al \xi^\mu) f^\al + \xi^\al D_\al f^\mu$.
An action in GR with a Lagrangian ${\cal L} (g_{\mu\nu},f^\nu)$
depending on the metric and matter fields is then given \mbox{generically as}
\beq
S = \int \sqrt{-g} \, d^4x \, {\cal L} (g_{\mu\nu},f^\nu) ,
\label{SGR}
\eeq
where $g$ is the determinant of the metric.
When ${\cal L}$ is a scalar, such an action in GR is invariant under local diffeomorphism
gauge transformations, obeying $\de S = 0$.

In GR, infinitesimal general coordinate transformations,
$x^\mu \rightarrow x^{\mu^\prime} (x) = x^\mu - \xi^\mu$,
are often called diffeomorphisms as well,
though technically they are the passive versions of the gauge-like transformations
(with the direction of $\xi^\mu$ reversed).
Nonetheless, changes of vector and tensor fields under these
transformations at a given coordinate $x^\mu$ are also given by their Lie derivatives.
Since a covariant theory will have an action that is unchanged under general
coordinate transformations, $\de S = 0$ holds as well under these~transformations.

At the level of effective field theory, 
diffeomorphisms are broken when a fixed background tensor,
denoted generically here as $\bar k_\ch$, appears in the Lagrangian.
The background $\bar k_\ch$ is a pre-specified field that in general has spacetime dependence, 
but it does not respond dynamically when it interacts with other fields in the theory.
As a fixed background, $\bar k_\ch$ does not transform under infinitesimal diffeomorphisms.
Instead, it obeys $\bar k_\ch \rightarrow \bar k_\ch$. 
Hence, an~action containing the metric, dynamical matter fields, and~a fixed background
may not be invariant, depending on how the symmetry is broken.
If the symmetry breaking is spontaneous,
then the background $\bar k_\ch$ is a vacuum expectation value of a fully dynamical field $k_\ch$.
In this case, $\bar k_\ch$ is accompanied in the action by NG and Higgs-like excitations,
and as a result the action remains invariant under diffeomorphisms.  
With spontaneous diffeomorphism breaking,
$\bar k_\ch$ is still a dynamical field, 
since it is a solution of the vacuum equations of motion.
In contrast, explicit diffeomorphism breaking occurs when the fixed background
is nondynamical.
In this case, $\bar k_\ch$ is not a solution of the equations of motion.
Instead, it is a pre-determined background that is inserted into the action,
sometimes referred to as an object with prior geometry.
With such a background tensor included,
the action takes the form 
\vspace{2pt} 
\beq
S = \int \sqrt{-g} \, d^4x \, {\cal L} (g_{\mu\nu},f^\nu,\bar k_\ch) ,
\label{Sbreaking}
\eeq
and $\de S \ne 0$ under~diffeomorphisms.

To include fermion fields in a gravity theory, a~vierbein formalism is used. 
Effectively, the~vierbein, $\vb \mu a$, gives tensor fields their components, 
at each spacetime point, with~respect to a local Lorentz basis,
where Latin letters are used to denote the local indices.
For the metric,
\beq
g_{\mu\nu} = \vb \mu a \vb \nu b \et_{ab} ,
\label{gvb}
\eeq
can be viewed as a defining relation for the vierbein,
where in this case its local components are those of the Minkowski metric $\et_{ab}$.
A generic matter field $f^\mu$ can be rewritten as $f^\mu = \ivb \mu a f^a$,
where $f^a$ are its components with respect to the local Lorentz frame,
and $\ivb \mu a$ is the inverse vierbein obeying $\ivb \mu a \vb \mu b = \de^a_b$
and $\ivb \mu a \vb \nu a = \de^\mu_\nu$.
The Lagrangian for a fermion field $\ps$, 
in the absence of diffeomorphism breaking,
can then be written as
$e {\cal L} = e i \ivb \mu a \bar \ps \ga^a D_\mu \ps$,
where $e$ is the determinant of the vierbein and $\ga^a$ is a Dirac matrix.
The covariant derivative $D_\mu$ acting on a fermion field is
$D_\mu \ps = \partial_\mu \ps + \fr 1 4 i \nsc \mu ab \si_{ab} \ps$,
where $\si_{ab} = \ga_a \ga_b - \ga_b \ga_a$
and $\nsc \mu ab$ is the spin connection.  
In addition to diffeomorphism invariance,
local Lorentz invariance becomes a second fundamental spacetime symmetry
in gravity theories when a vierbein formalism is introduced,
where the fermion fields and local tensor fields transform as
representations of the local Lorentz~group.  

When a fixed nondynamical background is included in a gravity theory
in a vierbein description, 
its components in both the spacetime frame and with respect to the
local Lorentz basis remain fixed under diffeomorphisms and local
Lorentz transformations.
For example,  $\bar k_\ch \rightarrow \bar k_\ch$ and
$\bar k_a \rightarrow \bar k_a$ under both types of spacetime transformations.
With both of these components being nondynamical and remaining fixed, 
they cannot be connected using the physical vierbein, 
e.g., $\bar k_\ch \ne \vb \ch a \bar k_a$.
Instead, they are related by a nondynamical vierbein denoted here as $\barvb \ch a$,
which does not transform under either diffeomorphisms or local Lorentz transformations.
Depending on whether $\bar k_\ch$, $\bar k_a$ or $\barvb \ch a$ appears
in a gravitational action,
either or both of local diffeomorphism invariance and local Lorentz invariance
can be explicitly~broken.

%%%%%%%%%%%%%%%%%%%%%%%%%%%%%%%%%%%%%%%%%%
\section{Consistency Conditions with Explicit~Breaking}\label{sec3}

Consider a gravity theory that contains a nondynamical background
$\bar k_\ch$ that explicitly breaks diffeomorphism invariance,
where $\ch$ generically denotes an arbitrary number of spacetime indices that
can be covariant, contravariant, or~mixed.
The action is
\beq
S = \int \sqrt{-g}\, d^4x  \left[ \fr 1 {16 \pi G} R + {\cal L} (g_{\mu\nu},f^\nu,\bar k_\ch) \right] ,
\label{Sbreaking2}
\eeq
where an Einstein--Hilbert term is used for the pure-gravity~sector.

Since $\bar k_\ch$ is not dynamical, it does not have equations of motion,
and its Euler--Lagrange expression does not need to vanish,
\beq
\fr {\de S} {\de \bar k_\ch} \ne 0.
\label{kEL}
\eeq

In contrast,
the metric $g_{\mu\nu}$ and matter fields $f^\nu$ are dynamical, 
and their equations of motion hold on shell.
For the metric,
its equations of motion yield the Einstein equations,
\beq 
G^{\mu\nu} = 8 \pi G T^{\mu\nu},
\label{EinsteinEq}
\eeq
where $G^{\mu\nu}$ is the Einstein tensor,
and $T^{\mu\nu}$ is the energy-momentum tensor that depends on the matter fields and
the background $\bar k_\ch$.
Since the contracted Bianchi identities require that $D_\mu G^{\mu\nu} = 0$ holds off shell,
consistency with the Einstein equations requires that $D_\mu T^{\mu\nu} = 0$ must hold
when the metric and matter fields are on shell.
However, the~fact that the background $\bar k_\ch$ appears in $T^{\mu\nu}$
and does not itself obey equations of motion,
calls into question whether consistency with the condition of energy-momentum
conservation, $D_\mu T^{\mu\nu} = 0$, can~hold.

General conditions that must hold for consistency in a theory with explicit
diffeomorphism breaking can be obtained as mathematical Noether identities
associated with general coordinate invariance.
Any physical theory should be independent of the choice of coordinates,
which requires that the action must be unchanged under a general coordinate transformation.
Although the backgrounds remain fixed under diffeomorphisms,
they do transform as tensors under observer-general coordinate transformations.
As noted above, 
when an infinitesimal general coordinate transformation 
$x^\mu \rightarrow x^{\mu^\prime} (x) = x^\mu - \xi^\mu$ 
is performed, vector and tensor fields transform with changes given
by their Lie derivatives.
This includes any fixed background tensors that are included in the action.
Thus, $\bar k_\ch \rightarrow \bar k_\ch + {\cal L}_\xi \bar k_\ch$ under infinitesimal 
general coordinate transformations.
With the action remaining unchanged under these transformations, obeying $\de S = 0$,
an off-shell identity can be found,
which relates the Euler--Lagrange expressions for
the metric, matter field, and~the background.
Putting the matter field on shell imposes that 
\beq
\fr {\de S} {\de f^\nu} = 0 .
\label{fEL}
\eeq

Combining this with the contracted Bianchi identity results in
four equations that must hold as a result of general coordinate invariance:
\beq
D_\mu T^{\mu}_{\pt{\mu} \nu} = 
- D_\ch (\fr {\de S} {\de \bar k_\ch} \bar k_\nu) + (D_\nu \bar k_\ch) \fr {\de S} {\de \bar k_\ch}  .
\label{Neother1}
\eeq

Here, for~simplicity, the~identity is written out for the case of a background tensor $\bar k_\ch$ 
that is a covariant vector.
For other types of tensors,
the Lie derivatives will differ and a different number of terms with different indices can appear,
with the corresponding Euler--Lagrange expression appearing in each term.
However, for~the purpose of pointing out the important features, a vector background~suffices.

Consistency with covariant energy-momentum conservation evidently requires 
that the right-hand side of \rf{Neother1} must vanish.
However, the~fact that the background does not satisfy Euler--Lagrange equations,
as indicated in Equation \rf{kEL},
makes this appear~problematic.

Nonetheless, the~right-hand side of \rf{Neother1} can vanish quite generally.
This is because, with explicit diffeomorphism breaking,
four local symmetries are lost, which results in the theory
having four additional degrees of freedom compared to a theory
with unbroken symmetry.  
These extra degrees of freedom become additional metric components,
which can take values that make $D_\mu T^{\mu\nu} = 0$ hold in \rf{Neother1}.   
Note that the metric will in general appear on the right-hand side of \rf{Neother1},
because the covariant derivatives depend on it.
However, if~the theory is limited in some way that does not permit a sufficient number of
degrees of freedom to appear,
then imposing $D_\mu T^{\mu\nu} = 0$ will put constraints on the background
or on the geometry,
resulting in a theory that is either inconsistent or uninteresting.
Examples of when this occurs include theories where an ansatz form of the metric
is used that does not have a sufficient number of degrees of freedom to satisfy
the four equations in \rf{Neother1}.
Alternatively, if~a linear perturbative treatment is used that suppresses the appearance of the
metric in nonlinear terms in \rf{Neother1},
the result again can be that the theory is inconsistent in such a~limit.

A St\"uckelberg approach is often used in theories with explicit diffeomorphism breaking to
reveal the extra degrees of freedom that appear.
This approach consists of introducing four dynamical scalar fields $\ph^A$ with $A=0,1,2,3$
and rewriting the background tensor in terms of these~\cite{ags03}.
For example, the~background $\bar k_{\mu\nu}$ used in massive gravity is \mbox{replaced by}
\beq
\bar k_{\mu\nu} \rightarrow (\partial_\mu \ph^A) (\partial_\nu \ph^B) \bar k_{AB} (\ph) ,
\label{kStuck}
\eeq
wherever it appears in the action.
This introduces four extra degrees of freedom in the form of the scalars $\ph^A$,
but it also restores the four diffeomorphism symmetries.
The net result is that there is no change in the number of degrees of freedom
compared to the original theory with explicit breaking.
For infinitesimal excitations,
the scalar can be expanded as $\ph^A = \de_\mu^A (x^\mu + \pi^\mu)$,
where $x^\mu$ are the coordinates and $\pi^\mu$ are the St\"uckelberg excitations.  
As can be seen when $\pi^\mu \rightarrow 0$,
the St\"uckelberg expression for $\bar k_{\mu\nu} $ in \rf{kStuck} reduces to
the original nondynamical tensor,
which shows that the explicit breaking theory is equivalent to a gauge-fixed
form of the St\"uckelberg modification.
However, if~the excitations are nonzero,
the expression for the background to first order in the excitations becomes
\beq
\bar k_{\mu\nu}  \rightarrow \bar k_{\mu\nu}  + (\partial_\mu \pi^\al) \bar k_{\al\nu}  (x) 
+ (\partial_\nu \pi^\al) \bar k_{\mu\al} + \pi^\al \partial_\al \bar k_{\mu\nu}  (x) .
\label{kStuck2}
\eeq

This expression can be recognized as the Lie derivative acting on $\bar k_{\mu\nu}$,
with $\pi^\mu$ playing the role of the transformation vector.
In a theory with spontaneous diffeomorphism breaking,
the excitations $\pi^\mu$ generate the broken symmetry transformations and
are therefore the NG excitations.
What the St\"uckelberg approach does in a theory with explicit breaking is
that it adds in the same NG modes as in the corresponding theory with
spontaneous breaking.
It also restores diffeomorphism invariance.  
However, note that there are no Higgs-like excitations generated in
the St\"uckelberg approach,
which distinguishes it from the case of spontaneous breaking.  
It is also the case that with explicit breaking, $\bar k_{\mu\nu}$ is still
not a dynamical solution,
whereas with spontaneous breaking, $\bar k_{\mu\nu}$ is a vacuum
solution for a dynamical tensor~field.

As a further example, the~St\"uckelberg approach can be used in the action in \rf{Sbreaking2}
involving a vector background $\bar k_\ch$.
In this case, the~replacement is $\bar k_\ch \rightarrow (\partial_\ch \ph^A) \bar k_A (\ph)$.
Since the scalars $\ph^A$ are dynamical in the St\"uckelberg approach,
the action in \rf{Sbreaking2} can be varied with respect to these fields to
obtain their equations of motion.
If the diffeomorphism gauge freedom is then fixed by setting $\ph^A = \de_\mu^A x^\mu$,
which sets the NG modes to zero,
the resulting equation of motion is
\beq
- D_\ch (\fr {\de S} {\de \bar k_\ch} \bar k_\nu) + (D_\nu \bar k_\ch) \fr {\de S} {\de \bar k_\ch}  = 0 .
\label{phStuck}
\eeq

Notice that this is the same expression as on the right-hand side of the identity in \rf{Neother1}.
Thus, when the equations of motion for the St\"uckelberg scalars are satisfied,
the identity in \rf{Neother1} sets $D_\mu T^{\mu\nu} = 0$.
However, there is no guarantee that the St\"uckelberg equations of motion in \rf{phStuck}
must hold.
Since the St\"uckelberg scalars themselves have been gauge fixed so that they equal
the coordinates,
it is the metric field, including four extra components that have been left unfixed by the gauge choice,
that must take values that \mbox{satisfy \rf{phStuck}}.
In general, this can occur,
but in models with an ansatz choice of metric or where a perturbative expansion is used,
it may not be possible for the metric to satisfy these equations.
Therefore, the~consistency issues can still persist when a St\"uckelberg approach is~used.

%%%%%%%%%%%%%%%%%%%%%%%%%%%%%%%%%%%%%%%%%%
\section{SME with Explicit Diffeomorphism~Breaking}\label{sec4}

The SME is a general theoretical framework for investigating diffeomorphism
and local Lorentz breaking. 
It is an effective field theory that maintains observer independence,
and it contains GR and the Standard Model (SM) as a subset of the theory.
The spacetime symmetry breaking tensors are referred to as SME coefficients,
and numerous bounds have been placed on them experimentally~\cite{aknr-tables}.  
A post-Newtonian limit of the SME has been developed~\cite{qbak06}
and matter--gravity interactions have been incorporated in a systematic way~\cite{akjt1,akjt2}.
These have been used to examine a number of experiments, including
lunar laser ranging tests~\cite{stubbs,angonin}, 
atom interferometry~\cite{mueller1,mueller2},
gravimeters~\cite{gravi},
short-range gravitational tests~\cite{bsl11,qbak15,jlak15,km17,Sh19},
analyses of baryon number asymmetry~\cite{GL06},
gyroscope precession~\cite{gyro},
perihelion and solar-spin tests~\cite{qbak06,Iorio},
orbital motion analyses~\cite{orb}, 
pulsar timing~\cite{pulsar1,pulsar2,pulsar3,pulsar4,pulsar5},
ring lasers~\cite{jt19},
and analyses of gravitational \v Cerenkov radiation~\cite{akjt3,ms18}.
Effects in gravitational radiation
have been investigated using a linearized version
of the SME~\cite{akmm16,akmm18,wei17}.

The same links as in GR between the Bianchi identities, dynamcs,
and energy-momentum conservation hold in the SME when
the background coefficients are assumed to be dynamical and arising as a result
of spontaneous spacetime symmetry breaking.
For this reason, the SME originally assumed that the symmetry breaking
was spontaneous and that the theory also contains NG and Higgs-like excitations
in addition to the background fields.
These small fluctuations play an important role in developing a useful
post-Newtonian limit of the SME and in incorporating matter--gravity couplings
in a systematic~way.

Field redefinitions also must be taken into account in gravity tests of
spacetime symmetry breaking,
because the SME coefficients are not all
physical or \mbox{independent~\cite{akgrav04,akjt1,akjt2,yb15,dcpm02}}.
For example, changes of coordinates or field redefinitions can be 
used to move some of the sensitivity to the 
symmetry breaking from one particle sector to another.
With gravity, ten components of the SME coefficients are unphysical,
which corresponds to having four coordinates and six local
Lorentz bases to choose.
As a result of these field redefinitions,
experiments with matter and gravity that are 
designed to test spacetime symmetry breaking
need to have sensitivity to more than one particle~sector.

For the purposes of this paper, it is sufficient to consider 
a fermion of mass $m$ and a photon field in gravity,
where the SME coefficients are limited to those in the minimal sector with
$s^{\mu\nu}$, $c_{\mu\nu}$, and~$(k_F)^{\ka\la\mu\nu}$.
With the Lorentz-invariant terms included,
the action is
\bea
S_{\rm SME} &\simeq& 
\int d^4x \, e \left[ \fr 1 {16 \pi G} [ R + s^{\mu\nu} R_{\mu\nu} ]
- \fr 1 4 F_{\ka\la} (g^{\ka\mu} g^{\la\nu} + (k_F)^{\ka\la\mu\nu}) F_{\mu\nu} \right.
\nonumber \\
&& \quad\quad\quad\quad\quad
\left.
+ i  \ivb \mu a \bar \ps ( \ga^a - c_{\al\be} \uvb \be a  \ivb \al b \ga^b) D_\mu \ps 
- m \bar \ps \ps \right] .
\label{SME}
\eea

The coefficients $s^{\mu\nu}$, $c_{\mu\nu}$, and~$(k_F)^{\ka\la\mu\nu}$
consist of background values and excitations,
\bea
s^{\mu\nu} &=& \bar s^{\mu\nu} + \tilde s^{\mu\nu} ,
\label{smunu1} 
\\
c_{\mu\nu} &=& \bar c_{\mu\nu} + \tilde c_{\mu\nu} ,
\label{cmunu1} 
\\
(k_F)^{\ka\la\mu\nu} &=& (\bar k_F)^{\ka\la\mu\nu} +  (\tilde k_F)^{\ka\la\mu\nu}  ,
\label{kappakalamunu1} 
\eea
where, in this notation, bars indicate the background values and tildes indicate
the excitations.
With spontaneous symmetry breaking,
the backgrounds are vacuum expectation values of dynamical tensors,
and the excitations include NG and Higgs-like modes.
In perturbative treatments,
the metric can also be expanded about a Minkowski background,
$g_{\mu\nu} = \et_{\mu\nu} + h_{\mu\nu}$,
which permits the construction of a post-Newtonian~framework.  

To consider the question of whether these terms in the SME can be used to
investigate gravity models with explicit diffeomorphism breaking,
a number of different features must be examined,
including how to maintain consistency between the Bianchi identities,
equations of motion, and~covariant energy-momentum conservation.  
When the symmetry breaking is explicit, 
the backgrounds are fixed and have no excitations, 
and hence the SME coefficients in \rf{SME} reduce to
\bea
s^{\mu\nu} &=& \bar s^{\mu\nu}  ,
\label{smunu2} 
\\
c_{\mu\nu} &=& \bar c_{\mu\nu}  ,
\label{cmunu2} 
\\
(k_F)^{\ka\la\mu\nu} &=& (\bar k_F)^{\ka\la\mu\nu} .
\label{kappakalamunu2} 
\eea

The backgrounds $\bar s^{\mu\nu}$, $\bar c_{\mu\nu}$, and~$(\bar k_F)^{\ka\la\mu\nu} $
are not solutions for dynamical tensors in the theory.
Instead, they are pre-determined fixed tensors put directly into the action.
While the backgrounds $\bar s^{\mu\nu}$, $\bar c_{\mu\nu}$, and~$(\bar k_F)^{\ka\la\mu\nu} $ 
have spacetime dependence, it is common in a first-order perturbative treatment about a 
flat Minkowski background to approximate them as~constant.

Although the backgrounds $\bar s^{\mu\nu}$, $\bar c_{\mu\nu}$, and~$(\bar k_F)^{\ka\la\mu\nu} $
have no excitations,
using a St\"uckelberg approach effectively introduces excitations.
Moreover, the~excitations take the same form as the NG modes in 
the corresponding and original version of the SME based on spontaneous breaking.
These can be denoted as $\tilde s^{\mu\nu}$, $\tilde c_{\mu\nu}$, and~$(\tilde k_F)^{\ka\la\mu\nu} $,
but unlike the SME based on spontaneous breaking,
they do not also include Higgs-like excitations.
However, insofar as maintaining the consistency of the model is concerned,
it suffices that the same NG modes appear as in the original SME with spontaneous breaking.
With the St\"uckelberg NG modes included,
$D_\mu T^{\mu\nu} = 0$ can hold while maintaining consistency with the Bianchi identities.  
Nonetheless, the~same caveats apply as described previously.
If enough degrees of freedom in the metric are suppressed,
either due to using a perturbative treatment or an ansatz solution,
then the geometry of the theory is severely restricted or the theory can be outright~inconsistent.  

An example of when the SME with explicit diffeomorphism breaking does not provide a useful framework 
is provided by the pure-gravity sector in a post-Newtonian limit,
where $\bar s^{\mu\nu}$ is the relevant SME coefficient.
When the St\"uckelberg approach is applied, this introduces the NG modes $\pi^\mu$. 
By gauge-fixing the St\"uckelberg scalars, 
the NG modes can be moved into the metric.
The additional excitations in $h_{\mu\nu}$ in a linearized approximation take the form 
$\partial_\mu \pi_\nu + \partial_\nu \pi_\mu$.
However, when these additional excitations are substituted into the linearized Ricci tensor
$R_{\mu\nu}$ in \rf{SME},
the NG modes drop out.  
This is because the linearized curvature is gauge-invariant,
and the excitations $\partial_\mu \pi_\nu + \partial_\nu \pi_\mu$ have a
pure-gauge form.
Without the NG modes,
it is not possible to satisfy the consistency conditions 
without putting severe restrictions on the geometry.
For example, with~$\bar s_{\mu\nu} \ne 0$, consistency of the theory requires that 
$\partial_\la R_{\mu\nu} = 0$ must hold in a post-Newtonian limit,
and such a condition placed on the curvature does not give a workable~model.

However, if~matter--gravity couplings are added,
the situation changes.
In this case, all three of the coefficients
$\bar s^{\mu\nu}$, $\bar c_{\mu\nu}$, and~$(\bar k_F)^{\ka\la\mu\nu} $ are included in 
the action in \rf{SME}.
Moreover, the~NG modes do not drop out in this case, 
since they couple to more than just the linearized curvature.
With matter--gravity couplings included, 
the SME can be used when there is explicit diffeomorphism breaking.
This follows because the same methodology used in the case of spontaneous breaking
is applicable with explicit breaking using a St\"uckelberg approach.
The St\"uckelberg method introduces the NG modes, but~not the Higgs-like modes.
However, in~the investigation in~\cite{akjt2},
it was assumed that the Higgs-like modes were either frozen out or
have negligible excitations.
Thus, the~same perturbative treatment can be applied
to the SME with explicit breaking as was worked out for the case
of spontaneous breaking.
In this method,
the symmetry properties allow the NG modes to be eliminated in terms
of the gravitational excitations and background SME coefficients.
The dominant signals of spacetime symmetry breaking can then
be extracted regardless of the details of the underlying~theory.

With all of the SME coefficients
$\bar s^{\mu\nu}$, $\bar c_{\mu\nu}$, and~$(\bar k_F)^{\ka\la\mu\nu} $ 
included in the action,
ten of them can be removed using field redefinitions.
Because of this,
it is important that experiments testing for spacetime symmetry
breaking in gravity have sensitivity to at least two of these sectors.
Otherwise, a~test in only one sector could involve a set
of coefficients that are unphysical.
With sensitivity to more than one set of SME coefficients,
experiments can obtain bounds assuming that one of the
sets has vanishing coefficients while making measurements
on an independent set.
Alternatively, bounds can be placed on combinations of
coefficients from two different particle sectors.
An example of the latter type is atom interferometry tests,
which are sensitive to both the gravity and photons.
The relevant coefficients in this case are $\bar s^{\mu\nu}$
and the symmetric trace $(\bar k_F)^{\al\mu\pt{\al}\nu}_{\pt{\al\mu}\al}$.
For comparisons across different experiments,
a Sun-centered celestial equatorial frame is used.
The components of the SME coefficients are obtained with
respect to this frame,
where indices are labeled using letters $JK\cdots$ for the spatial directions.
The combination of gravity and electromagnetic parameters that are bounded in 
these experiments are denoted as $\si^{JK}$,
and measured bounds of order $10^{-9}$ have been obtained for~them.

Ultimately, the~full consistency of a gravity theory with explicit diffeomorphism
breaking requires that the additional metric modes satisfy the identities stemming 
from general coordinate invariance for both the matter and gravity sectors at the same time.
For the gravity sector, this might involve nonlinear terms in a perturbative approach.
However, to~first order the results shown here for the SME with explicit breaking
indicate that the leading-order sensitivity to spacetime symmetry breaking
will be in the matter--gravity sector, 
since the pure-gravity sector does not have a useful first-order post-Newtonian~limit.

%%%%%%%%%%%%%%%%%%%%%%%%%%%%%%%%%%%%%%%%%%
\section{Examples with Matter~Couplings}\label{sec5}

Two examples of gravity theories with explicit diffeomorphism breaking
are examined in this section.
They are massive gravity and a model based on the IR limit of Ho\v rava gravity.
In massive gravity, a~fixed background tensor explicitly breaks diffeomorphism invariance,
while in Ho\v rava gravity, diffeomorphism invariance is broken by 
introducing a physical distinction between time and space in the form of a preferred foliation.
In both of these models, subject to some assumptions that must hold, 
matter--gravity interactions can appear that can be matched to SME terms in \rf{SME},
and experimental bounds can be placed on the relevant~couplings.

It is important to keep in mind that both massive gravity and Ho\v rava gravity
have experimental constraints in the pure-gravity sector and that other consistency
conditions involving stability, strong coupling in the gravity sector, black holes,
and cosmology must be addressed as well.
In fact, different formulations or modifications of the original versions of these theories 
have been proposed to address these other types of issues.
(See, for~example, the~reviews in~\cite{mg1,mg2,Muk10,Sot11,bps1,Wang17}.)
However, the~purpose of this paper is to examine how the SME can be applied,
and in this context, the matter--gravity couplings that are introduced in the following
subsections are independent of the original couplings in the pure-gravity sector,
and therefore any bounds that are obtained for these new couplings will be
independent of previously obtained bounds in the pure-gravity~sector.

%%%%%%%%%%%%%%%%%%%%%%%%%%%%%%%%%%%%%%%%%%
\subsection{Massive~Gravity}

A ghost-free massive gravity theory has been found by de Rham, Gabadadze, and~Tolley (dRGT)
\cite{mg1,mg2}.
To form mass terms for the metric,
a symmetric background tensor, denoted here as $\bar k_{\mu\nu}$,
is coupled to the metric in a nonlinear potential,
where the form of this potential eliminates the ghost mode.
As a result of introducing $\bar k_{\mu\nu}$, 
the dRGT Lagrangian explicitly breaks diffeomorphism invariance.
The dRGT action can be divided into a gravitational sector and a matter sector,
\beq
S_{\rm dRGT} =  S_{\rm grav} + S_{\rm matter} ,
\label{Smg}
\eeq
and either a metric or vierbein formalism can be~used.

In a metric description,
the action in the gravity sector has the form
\beq
S_{\rm grav} = \fr 1 {16 \pi G} \int d^4x \sqrt{-g} ( R - \fr {\mu^2} 4 {\cal U}({\mathbb X}) ) ,
\label{Smggrav}
\eeq
where $\mu$ is the graviton mass
and ${\cal U}({\mathbb X})$ is the potential that provides a mass term for the graviton. 
The potential is formed in terms of square roots ${\mathbb X}^\mu_{\pt{\mu}\nu}$,
which are defined as
\beq
{\mathbb X}^\mu_{\pt{\mu}\nu} = \sqrt{g^{\mu\al} \bar k_{\al\nu}} = \left( \sqrt{g^{-1} \bar k} \right)^\mu_{\pt{\mu}\nu} ,
\label{gasqrt1}
\eeq
and which obey
${\mathbb X}^\mu_{\pt{\mu}\al} {\mathbb X}^\al_{\pt{\mu}\nu} = g^{\mu\al} \bar k_{\al\nu}$.
In a vierbein description,
a background vierbein, $\bar v_{\mu}^{\pt{\mu}a}$, is introduced,
which is defined by $\bar k_{\mu\nu} = \bar v_{\mu}^{\pt{\mu}a} \bar v_{\nu}^{\pt{\mu}b} \et_{ab}$
The potential ${\cal U}$ can then be written as a sum of elementary symmetric polynomials
formed from products and traces of matrices 
\beq
\ga^\mu_{\pt{\mu}\nu} = \ivb \mu a \bar v_\nu^{\pt{\nu}a} .
\label{gasqrt}
\eeq

If the vierbein and background vierbein obey a symmetry condition,
\beq
\ivb \mu a  \bar v_{\mu b} = \ivb \mu b \bar v_{\mu a} ,
\label{symv}
\eeq
then the pure-gravity metric and vierbein descriptions are equivalent,
and ${\mathbb X}^\mu_{\pt{\mu}\nu} = \ga^\mu_{\pt{\mu}\nu}$
\cite{hr12,dmz}.

Couplings of matter to both the metric $g_{\mu\nu}$ and the background $\bar k_{\mu\nu}$
become likely when quantum corrections are included.
To avoid reintroducing the ghost,
the matter fields couple with an effective metric
$g^{\rm (eff)}_{\mu\nu}$ that is formed out of both the metric and the background field~\cite{matterdRGT1,matterdRGT2,NM15},
while the gravity sector remains unchanged.
The form of the effective metric is
\beq
g^{\rm (eff)}_{\mu\nu} = \al^2 g_{\mu\nu} + 2 \al \be \ga_{\mu\nu} + \be^2 \bar k_{\mu\nu} ,
\label{geff2}
\eeq
where $\ga_{\mu\nu} = g_{\mu\si} \ga^\si_{\pt{\mu}\nu}$
and $\al$ and $\be$ are constants.
Note that with lowered indices, the~square root matrix is symmetric, 
obeying $\ga_{\mu\nu} = \ga_{\nu\mu}$.
With $\be \ne 0$, matter couplings to $g^{\rm (eff)}_{\mu\nu}$ break local Lorentz symmetry
and diffeomorphisms.
However, since Lorentz breaking is known to be small,
to good approximation $\al \simeq 1$ while $\be \ll 1$,
and the effective metric therefore has the form
\beq
g^{\rm (eff)}_{\mu\nu} \simeq g_{\mu\nu} + 2 \be \ga_{\mu\nu} .
\label{geff3}
\eeq

It follows from this that the effective vierbein is 
\beq
{e^{\rm (eff)}_{\,\,\,\,\,\,\,\,\, \mu}}^{a} \simeq  \vb \mu a  + \be \vb \al a \ga^\al_{\pt{\al}\mu}  .
\label{eeffexpans}
\eeq

The matter terms in the action have the form
\beq
S_{\rm matter} = \int d^4x \, \sqrt{-g_{\rm (eff)}} \, {\cal L}_{\rm matter} (g^{\rm (eff)}_{\mu\nu}, f^\nu)  ,
\label{SdRGTmattermetric}
\eeq
or if fermions are included, it is given as
\beq
S_{\rm matter} = \int d^4x \, e_{\rm (eff)} \, {\cal L}_{\rm matter} ({e^{\rm (eff)}_{\,\,\,\,\,\,\,\, \mu}}^{a}, f^\nu,\ps) .
\label{SdRGTmattervierbein}
\eeq

Here, $f^\nu$ represents a tensor matter field, while $\ps$ denotes a fermion field.
These terms can be matched to the SME using
field redefinitions that change the effective metric back to the physical metric.
These fields' redefinitions also result in the introduction of SME~coefficients.

As an example, consider interactions in massive gravity 
involving photons with a fermion field.  
The action in this case is given as
\begingroup\makeatletter\def\f@size{9.5}\check@mathfonts
\def\maketag@@@#1{\hbox{\m@th\normalsize\normalfont#1}}%
\bea
S_{\rm dRGT} = \fr 1 {16 \pi G} \int d^4x \sqrt{-g} ( R - \fr {\mu^2} 4 {\cal U}({\mathbb X}) )
+ \int d^4x \, \sqrt{-g^{\rm (eff)}} \left(  - \fr 1 4 F_{\ka\la} \, g^{{\rm (eff)} \ka\mu} g^{{\rm (eff)} \la\nu}F_{\mu\nu} \right)
\nonumber \\
+ \int d^4x \, e^{\rm (eff)} [ i  {e^{\rm (eff)}}^\mu_{\,\,\, a}  \bar \ps \ga^a D^{\rm (eff)}_\mu \ps 
- m \bar \ps \ps ] ,
\quad\quad\quad\quad\quad\quad\quad
\label{SdRGT2}
\eea
\endgroup
where $m$ is the fermion mass.
Here, the~gravity sector depends on the physical metric $g_{\mu\nu}$
or vierbein $\vb \mu a$,
while the matter terms and $D^{\rm (eff)}_\mu$ depend on the effective metric or vierbein.
Field redefinitions can be used on $g^{{\rm (eff)} \mu\nu}$ 
and $e^{{\rm (eff)}\mu}_{\quad\quad a}$,
which at leading order reduce them to the
physical metric and vierbein,
while introducing SME coefficients defined in terms of 
the matrices $\ga_{\mu\nu}$.
In this way,
the massive gravity action $S_{\rm dRGT} $
can be expressed entirely in terms of the physical metric or vierbein,
but with additional interactions involving SME coefficients.
The result is 
\begingroup\makeatletter\def\f@size{9.3}\check@mathfonts
\def\maketag@@@#1{\hbox{\m@th\normalsize\normalfont#1}}%
\bea
S_{\rm dRGT} \simeq \fr 1 {16 \pi G} \int d^4x \sqrt{-g} ( R - \fr {\mu^2} 4 {\cal U}({\mathbb X}) )
+ \int d^4x \, \sqrt{-g} \left(  - \fr 1 4 F_{\ka\la} (g^{\ka\mu} g^{\la\nu} + (k_F)^{\ka\la\mu\nu}) F_{\mu\nu}  \right)
\nonumber \\
+ \int d^4x \, e [ i  e^\mu_{\,\,\, a}  \bar \ch ( \ga^a - c_{\al\be} \uvb \be a  \ivb \al b \ga^b) D_\mu \ch 
- m \bar \ch \ch ] ,
\quad\quad\quad\quad\quad
\label{SdRGT12}
\eea
\endgroup
where the relevant SME coefficients that arise are given as
\bea
c_{\mu\nu} \simeq \be (\ga_{\mu\nu} - \fr 1 4 \ga^\si_\si g_{\mu\nu}) ,
\label{cmunuSME}
\\
(k_F)^{\al}_{\pt{\al}\mu\al\nu} \simeq -2 \be (\ga_{\mu\nu} - \fr 1 4 \ga^\si_\si g_{\mu\nu}) .
\label{kSME}
\eea

Notice that $(k_F)^{\ka\la\mu\nu}$ has been partially traced and
that the fermion field $\ps$ has been rescaled and relabeled as $\ch$
so that the action remains in a standard Dirac~form.
Notice also that the Lorentz-violating couplings to the effective metric 
$g^{\rm (eff)}_{\mu\nu}$ have been replaced
by Lorentz-violating interactions with SME coefficients $(k_F)^{\al}_{\pt{\al}\mu\al\nu}$ and $c_{\mu\nu}$.
Because the field redefinitions were only on the matter sector,
the coefficients $u$ or $s^{\mu\nu}$ in \rf{SdRGT12} do not appear.
Thus, with~$\be \ne 0$,
it is not possible to make a further field redefinition to remove
$(k_F)^{\al}_{\pt{\al}\mu\al\nu}$ and $c_{\mu\nu}$ without generating 
$u$ and $s^{\mu\nu}$.

The potential term ${\cal U}$
is important in cosmology and gravitational radiation;
however, it has a negligible effect on matter--gravity tests, 
because the graviton mass is so small with
bounds on the order of $\mu \lsim 10^{-29}$ eV 
\cite{Nieto10}.
Thus, the~effects of $\mu$ can be ignored at leading order
in matter--gravity tests.
Furthermore, to~leading order, $g_{\mu\nu}$ and $\bar f_{\mu\nu}$ can be
approximated as Minkowski backgrounds,
in which case the contributions of $\ga^{\mu\nu}$ are also of order one.
With these approximations,
the spacetime symmetry breaking is determined by the parameter $\be$
in \rf{geff3} and \rf{eeffexpans},
and therefore $\be$ serves as a measure of the
symmetry-breaking matter couplings in massive~gravity.  

To obtain physical bounds, 
experiments that are sensitive to two sets of SME coefficients are needed.
In particular,
matter-interferometry experiments,
which have sensitivity in both the gravity 
and electromagnetic sectors of the SME~\cite{mueller1,mueller2},
are suitable for placing bounds at the level of $10^{-9}$ on
combinations of $s^{\mu\nu}$ and 
the symmetric trace $(k_F)^{\al\mu\pt{\al}\nu}_{\pt{\al\mu}\al}$.
As a result,
a bound of $\be \lsim 10^{-9}$ is obtained
on the matter--gravity couplings in dRGT massive~gravity. 

%%%%%%%%%%%%%%%%%%%%%%%%%%%%%%%%%%%%%%%%%%
\subsection{Ho\v rava-Based Model with a Preferred Spacetime~Foliation}

Ho\v rava gravity explicitly breaks diffeomorphism invariance
by introducing a physical preferred foliation of spacetime~\cite{ph09}.
Using coordinates $x^\mu = (t,x^i)$, with~$i=1,2,3$,
where constant values of $t$ distinguish the preferred foliations
and $x^i$ labels spatial points,
Ho\v rava gravity introduces an anisotropic scaling in
$t$ and $x^i$.
This permits terms to be added to the action
that maintain two time derivatives,
but which give a theory with power-counting renormalizability.
The preferred foliation explicitly breaks the full diffeomorphism
group to a subgroup consisting of time reparametrizations 
and three-dimensional spatial diffeomorphisms,
\beq
t \rightarrow t + \xi^0(t) ,
\label{tFPdiff}
\eeq
\beq
x^i \rightarrow x^i + \xi^i(t,x^j) ,
\label{xFPdiff}
\eeq

These are called foliation-preserving diffeomorphisms (FPdiffs).
Notice that under FPdiffs,
with $x^\mu \rightarrow x^\mu + \xi^\mu$,
the vector $\xi^\mu$ obeys $\partial_i \xi^0 = 0$.

The pure-gravity sector of Ho\v rava gravity has been explored extensively,
and a number of issues have been found that go beyond 
maintaining consistency with the Bianchi identities,
and as a result, a number of modified versions of Ho\v rava gravity
have been proposed~\cite{Muk10,Sot11,bps1,Wang17}.
In particular, the~role of the extra-metric modes in Ho\v rava gravity have been 
shown to play an important role in the pure-gravity sector~\cite{bps2,bps3,bps4}.
However, for~the purposes of this paper,
it is the matter--gravity couplings in a low-energy limit that 
can most readily be matched to the SME,
making the precise details of how the pure-gravity sector merges
with GR less critical to~consider.

If the low-energy or IR limit of Ho\v rava gravity is to merge with GR
and the SM, there should be a covariant formulation of Ho\v rava gravity in this limit.
There are different approaches that can be followed in constructing a 
covariant theory in the IR limit.
The approach examined here 
(see~\cite{no10,cog16,cas19} as an example of an alternative)
involves projecting covariant vectors and tensors
in four dimensions into the preferred foliation.
Such projections use a time-like normal vector
$n^\mu$, obeying $n_\mu n^\mu = -1$, 
which is orthogonal to the spatial foliations at each point.
These vectors are used in GR in a Hamiltonian formulation,
where the metric $g_{\mu\nu}$ is replaced by ADM%Define if appropriate
~variables
$(N,N^i,g_{ij})$,
where $N$ is the lapse, $N^i$ is the shift, and~$g_{ij}$ is the
three-dimensional spatial metric.
In GR, the~normal vectors $n^\mu$ can be defined with
respect to any foliation.
However, in~an IR limit of Ho\v rava gravity,
the vectors $n^\mu$ become fixed with respect to the
preferred foliation.
As a result, $n^\mu$ can transform under FPdiffs,
which maintain it as a normal vector,
but not under diffeomorphisms with $\partial_i \xi^0 \ne 0$,
since these would point it in a nonorthogonal direction.
A covariant theory constructed using $n^\mu$ therefore
explicitly breaks diffeomorphism invariance to FPdiffs,
and it can be used to model Ho\v rava gravity in the IR limit.
If a mapping of such a model to terms in the SME
can be found,
then existing experimental bounds on the SME terms can be used
to put limits on the corresponding couplings in the Ho\v rava-based~model.

The action in Ho\v rava gravity in the IR limit can be divided into two sectors,
\beq
S_{\rm Horava} = S_{\rm grav} + S_{\rm matter} ,
\label{Shorava}
\eeq
consisting of a gravity sector $S_{\rm grav}$ and a matter sector $S_{\rm matter}$,
where in principle both sectors can break diffeomorphism invariance to FPdiffs,
though with different coupling parameters.
The pure-gravity sector $S_{\rm grav}$ has been well studied,
including how it can be matched up with the gravity sector of the SME~\cite{qb21}.
It is defined using ADM variables for the metric,
the extrinsic curvature $K_{ij}$, its trace $K = g^{ij} K_{ij}$,
and the curvature tensor in the three-dimensional spatial slice.  
Its kinetic terms involve a combination $(K_{ij} K^{ij} - \la_g K^2)$,
where $\la_g$ is a running coupling constant.
Here, each of the products $K_{ij} K^{ij}$ and $K^2$ 
is separately a scalar under FPdiffs.
However, unless~$\la_g = 1$, their difference breaks diffeomorphism
invariance in four dimensions.  
There are also potential terms in $S_{\rm grav}$ consisting of 
contractions of higher-dimensional spatial components that make
Ho\v rava gravity power-counting renormalizable in the high-energy limit.
However, in~the IR limit, the~leading-order potential terms reduce to the three-dimensional 
curvature scalar $R^{\rm (3)} $ and a cosmological constant~term.

For Ho\v rava gravity to match with GR in the IR limit,
the running coupling $\la_g$ must approach 1,
in which case the kinetic and potential terms at leading order
reproduce the usual Einstein--Hilbert and $\La$ terms in GR
in ADM formalism.
However, with~small residual values of $(1- \la_g)$ in the IR limit,
there can still be an additional symmetry-breaking term.
With these assumptions,
an action modeling the gravity sector of Ho\v rava gravity in the IR limit 
can be written as
\beq
S_{\rm grav} \simeq \int \sqrt{-g} \, d^4 x \fr 1 {16 \pi G} (R - 2 \La)  
+ (1-\la_g) \fr 1 {16 \pi G} \int \sqrt{g^{\rm(3)}} N d^3x \, dt \, K^2  ,
\label{SgravIR}
\eeq
where $g^{\rm(3)}$ is the determinant of the three-dimensional metric.
In this way, $(1-\la_g) \ll 1$ becomes a parameter that measures 
the spacetime symmetry breaking in the gravity sector,
since it is the second term in $S_{\rm grav}$ that gives rise to the
spacetime symmetry breaking.  
It can be put in covariant form
using the timelike unit vector $n^\mu$ for the preferred foliation
to redefine the four-dimensional metric $g^{\mu\nu}$ as
\beq
g^{\mu\nu} = g^{ij} \de^\mu_i \de^\nu_j - n^\mu n^\nu ,
\label{gproj}
\eeq
where $g^{ij}$ is the three-dimensional inverse spatial metric.
The Kronecker delta functions appear as a result of using coordinates $(t,x^i)$,
where different $t$ values label the spacelike preferred foliations.
The extrinsic curvature can be written in terms of 
a covariant derivative of $n_\mu$ as $K_{ij} = \de^\mu_i \de^\nu_j D_\nu n_\mu$.

It is reasonable to assume that in the matter sector, $S_{\rm matter}$, 
diffeomorphism invariance is also explicitly broken to FPdiffs.  
This allows terms similar to those in the gravity sector,
where individually they are scalars under FPdiffs, 
but where they break 
four-dimensional diffeomorphisms.
In a low-energy limit that agrees with GR,
some subset of these terms would need to combine in such a way that
reproduces standard covariant matter-sector terms of dimension four or less.
Any residual diffeomorphism-breaking terms would have to have small
couplings to be consistent with~phenomenology.

%%%%%%%%%%%%%%%%%%%%%%%%%%%%%%%%%%%%%%%%%%
\subsubsection{Scalar and Vector~Fields}

As an example, consider a scalar field of mass $m$  in a gravitational theory.
In terms of ADM variables for the metric, 
the standard kinetic term for a scalar can be rewritten as
\beq
\half \partial_\mu \ph \partial^\mu \ph = - \fr 1 {2N^2} (\dot \ph - N^i \partial_i \ph)^2 + \half g^{ij} \partial_i \ph \, \partial_j \ph ,
\label{scalarADM}
\eeq 
where $\dot \ph = \partial_0 \ph$.
Together, the two terms on the right are diffeomorphism-invariant,
while individually, each term is separately invariant under FPdiffs.
Thus, in~a theory with a preferred foliation where FPdiffs is the fundamental symmetry,
a different linear combination of these terms can be used in the Lagrangian.
In this case, the~action for a scalar field in a Ho\v rava-based model in the IR limit is
\beq
S_{\rm scalar} \simeq 
\int \sqrt{g^{\rm(3)}} N \, d^3 x \, dt  \left[ c_1 (- \fr 1 {2N^2} (\dot \ph - N^i \partial_i \ph)^2 ) \right.
\left. + c_2 ( \half g^{ij} \partial_i \ph \, \partial_j \ph) - m^2 \ph^2 \right] ,
\label{Sscalar}
\eeq
where $c_1$ and $c_2$ are introduced as weighting parameters.
Making a projection with the normal vector, the~second term can be rewritten as
\beq
\half g^{ij} \partial_i \ph \partial_j\ph 
= \half (g^{\mu\nu} + n^\mu n^\nu) \partial_\mu \ph \partial_\nu \ph .
\label{projph}
\eeq

Then, by~combining \rf{scalarADM} and \rf{projph},
the action $S_{\rm scalar}$ becomes
\beq
S_{\rm scalar} \simeq 
\int \sqrt{-g} \, d^4 x \left[ c_2 (\half \partial_\mu \ph \partial^\mu \ph ) 
+ (c_2 - c_1) ( \half n^\mu n^\nu \partial_\mu \ph \, \partial_\nu \ph) - m^2 \ph^2  \right] .
\label{Sph2}
\eeq

A rescaling of the field $\ph$ and the mass $m$ can then be made,
which effectively sets $c_2 = 1$ and renames $c_1$ as $\la_\ph$,
giving the final form of the scalar action as
\beq
S_{\rm scalar} \simeq 
\int \sqrt{-g} \, d^4 x \left[ \half g^{\mu\nu} \partial_\mu \ph \, \partial_\nu \ph 
+ \half (1 - \la_\ph) n^\mu n^\nu \partial_\mu \ph \, \partial_\nu \ph - \half m^2 \ph^2 \right].
\label{Sph3}
\eeq

Thus, the~parameter $(1 - \la_\ph$) can be used to measure the extent of the diffeomorphism 
breaking in the matter sector,
similar to how $(1 - \la_g)$ is used in the gravity sector.
Since $\la_\ph$ can run as the energy scale changes,
just as $\la_g$ does in the gravity sector,
agreement with GR and the SM in the IR limit requires $(1 - \la_\ph) \ll 1$.

The Higgs sector of the SME has a coefficient $(k_{\ph\ph})^{\mu\nu}$ 
that can be matched with this coupling.
The correspondence is
\beq
(k_{\ph\ph})^{\mu\nu} = \half (1 - \la_\ph ) n^\nu n^\nu .
\label{kphph}
\eeq

However, with~three unbroken spatial diffeomorphisms,
the spatial components of $n^\mu$ can be gauged to zero
in any coordinate frame,
including the Sun-centered celestial equatorial frame.
With this choice,
only the purely timelike component of the SME coefficient
$(k_{\ph\ph})^{\mu\nu}$ remains nonzero 
and is given as
\beq
(k_{\ph\ph})^{00} = \fr 1 {2 N^2} (1 - \la_\ph ) .
\label{kph00}
\eeq

While bounds have been obtained for $(k_{\ph\ph})^{00}$ in the Higgs sector,
the experiments done to date have all assumed a Minkowski spacetime
and ignore gravitational effects.
For this reason, there are currently no experimental bounds
that can be placed on $(1 - \la_\ph )$ for the Higgs sector, 
since sensitivity to both gravity and the Higgs is~required.

A similar construction can be made for photon fields in the matter sector.
First, a~decomposition of the usual four-dimensional Lagrangian term for a 
massless photon can be made using the ADM formalism,
and for simplicity, the unbroken spatial diffeomorphisms can be used to set $N^i = 0$.
This reduces the usual photon Lagrangian term to
\beq
- \fr 1 4 F^{\mu\nu} F_{\mu\nu} = \fr 1 {2N^2} g^{ij} F_{0i} F_{0j}
- \fr 1 4 g^{ij} g^{kl}  F_{ik} F_{jl} .
\label{max2}
\eeq

Next,
since the decomposition splits the Lagrangian into
sectors that are individually scalars under FPdiffs,
a linear combination of the terms in \rf{max2} can be used in 
a gravity theory with a preferred foliation.
The resulting action in the IR limit is
\beq
S_{\rm photon} \simeq 
\int \sqrt{g} N \, d^4 x \, dt [c_1 \fr 1 {2N^2} g^{ij} F_{0i} F_{0j}
- c_2 \fr 1 4 g^{ij} g^{kl}  F_{ik} F_{jl} ] .
\label{Svec}
\eeq

Using projections defined with respect to the normal to the preferred foliation,
the following expressions are obtained,
\beq
g^{ij} F_{0i} F_{0j} = (g^{\mu\nu} + n^\mu n^\nu) F_{0\mu} F_{0\nu} ,
\label{gijFF}
\eeq
\beq
g^{ij} g^{kl} F_{ik} F_{jl} = (g^{\ka\la} g^{\mu\nu} 
+ 2 g^{\mu\nu} n^\ka n^\la) F_{\ka\mu} F_{\la\nu} .
\label{ggFF}
\eeq

Combining these with \rf{max2} allows \rf{Svec} to be rewritten in covariant form.
Then the parameters $c_1^{(\ga)}$ and $c_2^{(\ga)}$ in \rf{Svec} can be redefined,
respectively, as $c_1^{(\ga)} = \la_\ga$ and $c_2^{(\ga)} = 1$.
The action describing a massless vector in the IR limit of a Ho\v rava-based model 
then matches the form of a term in the photon sector of the SME,
\beq
S_{\rm photon} \simeq 
\int \sqrt{-g} \, d^4 x (- \fr 1 4 F^{\mu\nu} F_{\mu\nu} 
-  \fr 1 4 (k_F)^{\ka\la\mu\nu} F_{\ka\la} F_{\mu\nu}  ) ,
\label{Svec2}
\eeq
where the SME coefficient $(k_F)^{\ka\la\mu\nu} $ in this context
is given as
\beq
(k_F)^{\ka\la\mu\nu} = \half (1 - \la_\ga) [ g^{\ka\mu} n^\la n^\nu 
- g^{\ka\nu} n^\la n^\mu
- g^{\la\mu} n^\ka n^\nu + g^{\la\nu} n^\ka n^\mu ]  .
\label{kFHorava}
\eeq

Here, $n^\nu$ acts effectively as a partially fixed background
that does not transform under diffeomorphisms
with $\partial_i \xi^0 \ne 0$.
With the gauge fixing setting $n^\mu = (\fr 1 N,0)$,
the only nonzero SME components are
\beq
(k_F)^{0i0j} = \fr 1 {2N^2} (1 - \la_\ga) \, g^{ij} .
\label{kvalues}
\eeq

Thus, with~$|g^{ij}| \simeq 1$ and $N \simeq 1$,
gravity experiments with sensitivity to $(k_F)^{0i0j}$
can be used to place bounds on the parrameter $(1 - \la_\ga)$.
In this example,
there are experiments with sensitivity to both gravity and electromagnetism.
These are the atom interferometry tests,
which place bounds at a level of $10^{-9}$ on the quantities $\si^{JK}$  
in the Sun-centered celestial equatorial frame~\cite{mueller1,mueller2}.
With these experimental results, the~bounds on photon-gravity couplings in the
Ho\v rava-based model are estimated as
$ |1 - \la_\ga | \lsim 10^{-9} $.

%%%%%%%%%%%%%%%%%%%%%%%%%%%%%%%%%%%%%%%%%%
\subsubsection{Vierbein Description and~Fermions}

To include matter couplings to fermions,
a vierbein formalism can be used.
The vierbein $\vb \mu a$ provides a link between
tensor components in a local Lorentz frame with those in the spacetime frame.
It can be used on the normal vector, $n^\mu$, which is orthogonal to the preferred 
foliations in a Ho\v rava-based model, to find the corresponding components $n^a$
in the local frames.
Since there is freedom to choose and align the orientation of the local Lorentz bases,
a simplifying choice is to pick one that gives $n^a = (1,0,0,0)$ and $n_a = \et_{ab} n^b = (-1,0,0,0)$,
which together obey $n^a n_a = -1$.
Using these as projections,
the vierbein in four dimensions can be decomposed into components with time indices
and a three-dimensional spatial subblock.
The local indices can be distinguished as
$a = 0,A$ with $A=1,2,3$ labeling the local spatial components.
With these, a~three-dimensional spatial vierbein $\vb i A$ is defined 
such that it equals $\vb \mu a$ when $\mu = i$ and $a = A$.
The four-dimensional inverse vierbein can then be decomposed as
\beq
\ivb \mu a = \de^\mu_i \de^A_a \ivb i A - n_a n^\mu .
\label{vbproj}
\eeq
where $\ivb i A$ is the three-dimensional inverse of $\vb i A$.
When this is substituted into $g^{\mu\nu} = \ivb \mu a \ivb \nu b \et^{ab}$,
it reproduces \rf{gproj}.
Note also that $\ivb i A \ivb j B \et^{AB} = g^{ij}$.

In local Lorentz frames, $n^a$ remains orthogonal to the preferred spacelike foliation
in the same way that $n^\mu$ remains orthogonal to the preferred foliation
in the spacetime frame.  
For this reason, $n^a$ must stay fixed under local Lorentz boosts,
since otherwise its orthogonality to the spatial foliations cannot be maintained.
In this way, local Lorentz transformations are explicitly broken
when a vierbein description is used that introduces $n^a$,
similarly to how $n^\mu$ explicitly breaks diffeomorphisms.
However, spatial rotations in the local frame remain symmetrical,
which is similar to how FPdiffs remain symmetrical in the spacetime~description.

The usual relativistic kinetic Lagrangian term for a fermion $\ps$ with mass $m$
is given as
\beq
{\cal L}_\ps = i \ivb \mu a \bar \ps \ga^a D_\mu \ps .
\label{Lps}
\eeq
To obtain terms that are suitable for 
describing a fermion in a gravity
model based on the IR limit of Ho\v rava gravity,
this fully relativistic term can be subdivided into two parts,
\beq
{\cal L}_\ps = {\cal L}_1  + {\cal L}_2 ,
\label{L12}
\eeq
where each part, ${\cal L}_1$ and ${\cal L}_2$, 
is invariant individually under FPdiffs and local spatial rotations.
The contractions in the purely spatial subblock define ${\cal L}_2$ as
\beq
{\cal L}_2 =  i \ivb i A \bar \ps \ga^A D_i \ps ,
\label{LiA}
\eeq
while the terms that come from summing out the remaining components in $\mu$ and $a$ in \rf{Lps}
define ${\cal L}_1$,
which for simplicity is not written out here.
Using \rf{vbproj},
${\cal L}_2$ becomes
\bea
{\cal L}_2 &=& i(\ivb \mu a + n_a n^\mu) \bar \ps \ga^a D_\mu \ps
\\ \nonumber
&=& 
{\cal L}_\ps + i n_a n^\mu \bar \ps \ga^a D_\mu \ps .
\label{proj2}
\eea

Since ${\cal L}_1$ and ${\cal L}_2$ are each separately
scalars under FPdiffs and local spatial rotations,
a linear combination of these terms can be used
in a Ho\v rava-based model for a fermion,
\beq
{\cal L}_{\rm fermion} = c_1 {\cal L}_1  + c_2 {\cal L}_2 .
\label{LHorava}
\eeq 
This can be rewritten as
\bea
{\cal L}_{\rm fermion} &=& c_1 ({\cal L}_\ps - {\cal L}_2)  + c_2 {\cal L}_2 
\nonumber\\ 
&=& c_1 {\cal L}_\ps + (c_2 - c_1) {\cal L}_2 
\nonumber\\ 
&=& c_1 {\cal L}_\ps + (c_2 - c_1) ( {\cal L}_\ps + i n_a n^\mu \bar \ps \ga^a D_\mu \ps)
\\ \nonumber
&=& c_2 {\cal L}_\ps + (c_2 - c_1)  i n_a n^\mu \bar \ps \ga^a D_\mu \ps .
\label{Lhoravaeqs}
\eea
Lastly, rescaling so $c_2 =1$ and $c_1 = \la_\ps$ gives the final form,
\beq
{\cal L}_{\rm fermion} = {\cal L}_\ps + (1- \la_\ps)  i n_a n^\mu \bar \ps \ga^a D_\mu \ps .
\label{LHorav}
\eeq
Matching this to the SME term
\beq
{\cal L}_{\rm SME} = -i \ivb \mu a \bar \ps c_{\al\be} e^{\be a} \ivb \al b \ga^b D_\mu \ps
\label{fermionSME}
\eeq
identifies the SME parameter for the Ho\v rava-based model as
\beq
(1- \la_\ps)  n_a n^\mu = - \ivb \mu b c_{\al\be} \, e^{\be b} \ivb \al a ,
\label{calbe}
\eeq
or
\beq
c_{\mu\nu} = -(1- \la_\ps)  \vb \mu a n_a n_\nu .
\label{cmunu2}
\eeq

Experiments with sensitivity to matter--gravity couplings involving protons, neutrons, and~electrons
have been analyzed in~\cite{akjt2,MullerPRL13}.
They obtain bounds on the SME parameter $(\bar c)_{TT}$ for the time components
in the Sun-centered celestial equatorial frame in the range of $10^{-6} - 10^{-8}$,
depending on the type of experiment and assumptions concerning nuclear binding energies.
These can be used to estimate a bound on the coupling $(1- \la_\ps)$,
which limits it to the range $| 1- \la_\ps | \lsim 10^{-6} - 10^{-8}$.

%%%%%%%%%%%%%%%%%%%%%%%%%%%%%%%%%%%%%%%%%%
\section{Summary and~Conclusions}\label{sec6}

The question of whether the SME based on the process of spontaneous diffeomorphism
breaking can also be used to analyze models with explicit spacetime symmetry breaking
has been investigated.
The answer hinges on whether consistency between the Bianchi identities,
the equations of motion, and~covariant energy-momentum conservation
can be maintained.
Identities relating these can be obtained for theories that maintain general coordinate invariance,
and they reveal that, in general, consistency can be maintained as long as the additional
metric modes that come from the diffeomorphism breaking are not suppressed in
the consistency conditions.
In perturbative treatments, such as in a first-order post-Newtonian limit,
the additional metric excitations are suppressed,
and therefore there is no useful post-Newtonian limit.
However, when matter--gravity couplings are considered,
the result is different.
Using a St\"uckelberg approach,
the additional excitations in a theory with explicit breaking due to a fixed
background tensor are found to be the same as the NG excitations in a
corresponding theory with spontaneous breaking.
Therefore, the~same methodology can be applied in theories with
explicit diffeomorphism breaking,
and the SME can be used as a phenomenological framework for
making comparisons with~experiments.

Two examples of theories with explicit diffeomorphism breaking have been examined,
where possible matter--gravity couplings have been investigated using the SME.
In massive gravity,
experiments with sensitivity to gravitational and electromagnetic interactions 
place bounds on the order of $10^{-9}$ on possible photon couplings. 
Similarly, a~model based on the IR limit of Ho\v rava gravity includes matter--gravity
terms that can be compared to the SME.
In this case, 
sensitivity to photon couplings on the order of $10^{-9}$
and to fermion couplings in the range $10^{-6}$--$10^{-8}$
can be~attained.

%%%%%%%%%%%%%%%%%%%%%%%%%%%%%%%%%%%%%%%%%%

%%%%%%%%%%%%%%%%%%%%%%%%%%%%%%%%%%%%%%%%%%
\end{document}